\documentstyle[11pt]{article}
\def\div{{\,\rm div \,}}
\def\rank{{\,\rm rank \,}}
\def\so{{\,\rm so \,}}
\def\qaq{{\quad\mbox{at}\quad}}

\def\CC{{\,\rm C\,}}

\def\sym{{\,\rm sym \,}}

\def\ii{{\,\rm i \,}}

\def\+M{{\,\rm M^{n\times n}_+ \,}}
\def\tr{{\,\rm tr \,}}

\def\qfq{{\quad\mbox{for}\quad}}

\def\ii{{\,\rm i \,}}

\def\lam{\lambda}

\def\E{{\cal E}}

\def\X{{\cal X}}

\def\E{{\cal E}}

\def\N{{\cal N}}

\newfont{\Blackboard}{msbm10 scaled 1200}

\newfont{\roma}{cmr10 scaled 1200}

\def\<{{\langle}}
\def\>{{\rangle}}

\def\g{\gamma}

\def\var{\varphi}
\def\si{\sigma}

\def\a{\alpha}
\def\b{\beta}

\def\Om{\Omega}

\newtheorem{thm}{{}\hskip\parindent Theorem}[section]
\newtheorem{lem}{{}\hskip\parindent Lemma}[section]
\newtheorem{pro}{{}\hskip\parindent Proposition}[section]

\newtheorem{cor}{{}\hskip\parindent Corollary}[section]

\newtheorem{rem}{{}\hskip\parindent Remark}[section]

\def\pl{\partial}
\def\rw{\rightarrow}

\def\be{\begin{equation}}
\def\ee{\end{equation}}
\def\beq{\arraycolsep=1.5pt\begin{eqnarray}}
\def\eeq{\end{eqnarray}}

\def\R{I\!\!R}

\def\n{\vec{n}}
\large
\title{Optimal exponentials of thickness in Korn's inequalities  for parabolic and elliptic shells}
\date{}
\author{
Peng-Fei YAO\\[0.3cm]
Key Laboratory of  Systems and Control\\
Institute of Systems Science,
Academy of Mathematics and Systems Science\\
Chinese Academy of Sciences, Beijing 100190, P. R. China\\
School of Mathematical Sciences\\
University of Chinese Academy of Sciences, Beijing 100049, China\\
e-mail: pfyao@iss.ac.cn}

\begin{document}
\maketitle
 \footnote{This work is  supported by the National
Science Foundation of China, grants  no. 61473126 and no. 61573342, and Key Research Program of Frontier Sciences, CAS, no. QYZDJ-SSW-SYS011.}

\begin{quote}
\begin{small}
{\bf Abstract} \,\,\,We establish Korn's interpolation inequalities and the rigidity results of the strain tensor of the middle surface for the parabolic  and  elliptic shells and show that
the best constant in Korn's inequalities scales like $h^{3/2}$  for the parabolic shell and $h$ for the elliptic shell, removing the main assumption that the middle surface of the shell is given by one single principal coordinate in the literature and, in particular, including the closed elliptic shell.
\\[3mm]
\end{small}
\end{quote}

\setcounter{equation}{0}
\section{Introduction and Main Results}
\def\theequation{1.\arabic{equation}}
\hskip\parindent
Korn¡¯s inequalities have arisen in the investigation of the boundary value problem of linear
elastostatics, \cite{Ko,Ko1} and have been proven by different authors, e.g., \cite{Fr, Kond, Kond1, Koh, PaWe}.
Some generalized versions of the classical second Korn inequality have been recently proven
in \cite{BNPS, CDM, NPW, NPW1}. The optimal exponential of thickness in Korn's inequalities for thin shells represents the relationship between the rigidity and the thickness of a shell when the small deformations take place since  Korn's inequalities are linearized from the geometric rigidity inequalities under the small deformations (\cite{FrJaMu}). Thus it is the best Korn constant in the Korn inequality that is of central importance (e.g., \cite{CiOlTr,LeMu, Na, Na1, PaTo, PaTo1}). Moreover, it is  ingenious that the best Korn constant is subject to the Gaussian curvature. The one for the parabolic shell  scales like $h^{3/2}$ (\cite{GH,GH1}), for the hyperbolic shell, $h^{4/3}$ (\cite{Ha2}) and for the elliptic shell, $h$ (\cite{Ha2}). All those results were derived under the main assumption that the middle surface of the shell is given by a single principal coordinate system in order to carry out some necessary computation. This assumption is
\be S=\{\,{\bf r}(z,\theta)\,|\,(z,\theta)\in[1,1+l]\times[0,\theta_0]\,\},\label{as}\ee where the properties
$$\nabla_{\pl z}\n=\kappa_z\pl z,\quad\nabla_{\pl\theta}\n=\kappa_\theta\pl\theta\qfq p\in S $$ hold.

In the case of the parabolic or hyperbolic shell, a principal coordinate only exists locally (Proposition \ref{pr2.1}). There is even no such a local existence for the elliptic shell. However, the  assumption (\ref{as}) in \cite{GH,GH1, Ha2} can be removed if the Bochner technique is employed to perform  some necessary computation. The Bochner technique
provides us the great simplification in computation, for example, see \cite{Ho} or \cite{Yao2011}.  Here we remove the assumption (\ref{as}) to obtain that the optimal exponentials are $3/2$ and $1$ for the parabolic shell and  the elliptic shell, respectively. In particular,  the closed elliptic shell is included here.  The case of the hyperbolic shell is  treated in \cite{Yao2018} where we show that the optimal exponential is $4/3$ without the assumption (\ref{as}).

Let $M\subset\R^3$ be a $\CC^3$ surface with the induce metric $g$ and a normal field $\n.$ Let $S\subset M$ be an open bounded set with a regular boundary $\pl S.$ We consider a shell with thickness $h>0$
$$\Om=\{\,x+t\n(x)\,|\,x\in S,\,-h<t<h\,\}.$$
Let $\kappa$ be the Gaussian curvature of $M.$ We say that $\Om$ is parabolic if
\be\kappa(x)=0,\quad|\Pi(x)|>0\qfq x\in\overline{S},\label{1.2}\ee where $\Pi=\nabla\n$ is the second fundamental form of $M.$
If
\be\kappa(x)>0\qfq x\in\overline{S},\label{1.6}\ee
then $\Om$ is said to be elliptic.

Set
$$H_0^1(\Om,\R^3)=\{\,y\in H^1(\Om,\R^3)\,|\,y|_{\Sigma_0}=0\,\},$$ where
$$\Sigma_0=\{\,x+t\n(x)\,|\,x\in\pl S,\,\,|t|\leq h\,\}.$$
Here it can happen that $\pl S=\emptyset,$ for example, to a closed elliptic shell, for which $H_0^1(\Om,\R^3)=H^1(\Om,\R^3).$

All the norm $\|\cdot\|$ in this paper is that of $L^2(\Om),$ unless it is specified.

\begin{thm}\label{t1}\mbox{$($Korn's interpolation inequalities$)$} There are $C>0,$ $h_0>0,$ independent of $h>0,$ such that
\be \|\nabla y\|^2\leq C(\frac1h\|\<y,\n\>\|\|\sym\nabla y\|+\|y\|^2+\|\sym\nabla y\|^2)\label{1.1}\ee
for all $h\in(0,h_0)$ and $y\in H^1(\Om,\R^3)$ with $\<y,\n\>|_{\Sigma_0}=0$  where
$$\sym\nabla y=\frac12(\nabla y+\nabla^Ty).$$
\end{thm}

We have the following.
\begin{thm}\label{t2} Let $\Om$ be parabolic.
There are $C>0,$ $h_0>0,$ independent of $h>0,$ such that
\be\|\nabla y\|^2\leq\frac{C}{h^{3/2}}\|\sym\nabla y\|^2,\label{1.3}\ee
for all $h\in(0,h_0)$ and $y\in H_0^1(\Om,\R^3).$
\end{thm}

\begin{thm}\label{t3} Let $\Om$ be elliptic.
There are $C>0,$ $h_0>0,$ independent of $h>0,$ such that
\be\|\nabla y\|^2\leq\frac{C}{h}\|\sym\nabla y\|^2\label{1.7}\ee
for all $h\in(0,h_0)$ and $y\in H_0^1(\Om,\R^3).$
\end{thm}

In particular, we have

\begin{cor}\label{cor}If $\Om$ is a closed elliptic shell, then there is $C>0$ such that
\be\min_{A\in\so(3)}\|\nabla y-A\|^2\leq\frac{C}{h}\|\sym\nabla y\|^2\label{zui}\ee for any $y\in H^1(\Om,\R^3),$ where $\so(3)$ is the set of all $3\times3$ skew matrices.
\end{cor}

\begin{thm}\label{t4}
The exponentials  of the thickness  in $(\ref{1.3})$ and $(\ref{1.7})$-$(\ref{zui})$ are optimal, respectively, for the parabolic shell and the elliptic shell, respectively.
\end{thm}

\begin{rem} The interpolation inequality $(\ref{1.1})$ is given in $\cite{GH, GH1, Ha2}$ under the assumption $(\ref{as})$ and extended in $\cite{Ha}$ to the case that there is a local principal coordinate for each $p\in S.$ The inequalities $(\ref{1.3})$ and $(\ref{1.7})$ are given in $\cite{GH1,Ha}$ and $\cite{Ha2},$ respectively, under the assumption $(\ref{as}).$

\end{rem}

\setcounter{equation}{0}
\section{Proofs of Main Results}
\def\theequation{2.\arabic{equation}}
\subsection{Proof Theorem \ref{t1}}
\hskip\parindent Let $(M,g)$ be a Riemanniann manifold.  Let $T$ be a 2-order tensor field on $(M,g)$ and let $X$ be a vector field on $(M,g).$ We define the inner multiplication of $T$ with $X$ to be another vector field, denoted by $\ii(X)T,$ given by
$$\<\ii(X)T,Y\>=T(X,Y)\qfq Y\in M_p,\quad p\in M,\quad g=\<\cdot,\cdot\>.$$

For any $y\in H^1(\Om,\R^3),$ we decompose $y$ into
\be y(z)=W(x,t)+w(x,t)\n(x)\qfq z=x+t\n(x)\in\Om,\quad x\in S,\quad |t|<h,\label{2.1}\ee where $w=\<y,\n\>$ and $W(\cdot,t)$ is a vector field on $S$ for $|t|<h.$ It follows from (\ref{2.1})
that
\be \nabla_{\a+t\nabla\n\a} y=D_\a W+w\nabla_\a\n+[\a(w)-\Pi(W,\a)]\n\qfq \a\in S_x,\ee
\be\nabla_{\n}y=W_t(x,t)+w_t(x,t)\n(x)\qfq  x\in S,\quad|t|<h,\label{2.2}\ee
where $\nabla$ and $D$ are the covariant differentials of the dot metric in $\R^3$ and of the induced metric in $S,$ respectively, and
$W_t=\pl_tW$ and $w_t=\pl_tw.$ We need to deal with the relations between $\nabla$ and $D$ carefully.

By defining $\nabla\n\n=0,$ we introduce an 2-order tensor $p(y)$ on $\R^3_x$ by
\be p(y)(\tilde\a,\tilde\b)=\<\nabla_{\nabla\n\tilde\a}y,\tilde\b\>\qfq\tilde\a,\,\,\tilde\b\in\R^3.\label{2.4}\ee

We have
\begin{lem}\label{lem2.1} Let $y\in H^1(\Om,\R^3)$ be given in $(\ref{2.1}).$ Then
\be|\nabla y+tp(y)|^2=|DW+w\Pi|^2+|Dw-\ii(W)\Pi|^2+|W_t|^2+w_t^2,\label{2.5}\ee
\be|\sym\nabla y+t\sym p(y)|^2=|\Upsilon(y)|^2+\frac12|X(y)|^2+w_t^2,\label{2.6}\ee where
\be\Upsilon(y)=\sym DW+w\Pi,\quad X(y)=Dw-\ii(W)\Pi+W_t.\label{2.7}\ee
\end{lem}

{\bf Proof}\,\,\,Let $x\in S$ be given.
Let $\tau_1,$ $ \tau_2$ be an orthonormal basis of $S_x.$ Then $\tau_1,$ $\tau_2,$ and $\n(x)$ forms an orthonormbal basis of $\R^3_x.$ From (\ref{2.2}) and (\ref{2.2}), we have
\beq|\nabla y+tp(y)|^2&&=\sum_{ij=1}^2\<\nabla_{\tau_i+t\nabla\n\tau_i}y,\tau_j\>^2+\sum_{i=1}^2(\<\nabla_{\tau_i+t\nabla\n\tau_i}y,\n\>^2+\<\nabla_{\n} y,\tau_i\>^2)+\<\nabla_{\n} y,\n\>^2\nonumber\\
&&=|DW+w\Pi|^2+|Dw-\ii(W)\Pi|^2+|W_t|^2+w_t^2,\nonumber\eeq
\beq|\sym\nabla y+t\sym p(y)|^2&&=\sum_{ij=1}^2[\frac12(\<\nabla_{\tau_i+t\nabla\n\tau_i}y,\tau_j\>+\<\nabla_{\tau_j+t\nabla\n\tau_j}y,\tau_i\>)]^2\nonumber\\
&&\quad+\sum_{i=1}^2[\frac12(\<\nabla_{\tau_i+t\nabla\n\tau_i}y,\n\>+\<\nabla_{\n} y,\tau_i\>)]^2+\<\nabla_{\n} y,\n\>^2\nonumber\\
&&=|\sym DW+w\Pi|^2+\frac12|Dw-\ii(W)\Pi+W_t|^2+w_t^2.\nonumber\eeq \hfill$\Box$

\begin{rem} $\Upsilon(y)$ and $DX$ are called the strain tensor and the curvature tensor of the middle surface, respectively, see $\cite{Koi}.$
\end{rem}

We need the following lemma from \cite{Ha1}.

\begin{lem}\label{l2.3}$(\cite{Ha1})$ Assume $\lam\in(0,1],$ $0\leq a<b$ and $f$ : $[a, b] \rw R$ is absolutely continuous. Then
the inequality holds:
\be \int_{a+\lam(b-a)}^bf^2(t)dt\leq\frac{2}{\lam}\int_a^{a+\lam(b-a)}f^2(t)dt+4\int_a^b(b-t)^2f'^2(t)dt.\label{new2}\ee
\end{lem}

For $f\in L^2(\Om),$ we have
$$\int_\Om f^2(z)dz=\int_{-h}^h\int_Sf^2(x+t\n(x))(1+t\tr\Pi+t^2\kappa)dgdt,$$ where $\kappa$ is the Gaussian curvature. It follows that
\be(1-Ch)\int_{-h}^h\int_Sf^2(x+t\n)dgdt\leq\int_\Om f^2(z)dz\leq (1+Ch)\int_{-h}^h\int_Sf^2(x+t\n)dgdt.\label{2.10}\ee
In the sequel, we sometimes use the norm
\be\|f\|^2=\int_{-h}^h\int_Sf^2dgdt\qfq h\quad\mbox{small}\label{b3}\ee instead of the norm
$$\|f\|^2=\int_\Om f^2dz.$$

The next lemma is the key to our analysis that is the 3-dimensional version of \cite[Lemma 4.5]{Ha}. In the 2-dimensional case \cite[Lemma 4.5]{Ha}
establishes the inequality (\ref{2.12}) without the assumption (\ref{2.11}) below.

Now we consider the product manifold
$$\Om(h)=S\times(-h,h).$$ We denote by $\hat\nabla$ and $\hat\Delta$ the connection and the Laplacian of the product manifold $\Om(h),$ respectively. If $w$ is a function on $\Om(h),$ then
$$\hat\nabla w=Dw+w_t\pl t,\quad \hat\Delta w=\Delta_gw+w_{tt},$$ where $\Delta_g$ is the Laplacian of the induced metri $g.$

\begin{lem}\label{l2.4} There is a constant $C>0,$ independent of $h>0,$ such that any harmonic function $w$ on the product manifold $S\times(-h,h)$  with
\be w|_{\Sigma_0}=0\label{2.11}\ee fulfills the
inequality
\be\|\hat\nabla w\|^2\leq C(\frac1h\|w\|\|w_t\|+\|w_t\|^2).\label{2.12}\ee
\end{lem}

{\bf Proof}\,\,\,For given $t\in(0,h),$ we have
\beq\int_{-t}^t\int_S|\hat \nabla w|^2dgdt&&=\int_{-t}^t\int_S[-w(\Delta_gw+w_{tt})+\div_gwDw+(ww_t)_t]dgdt\nonumber\\
&&=\int_S[w(x,t)w_t(x,t)-w(x,-t)w_t(x,-t)]dg\nonumber\\
&&\leq\int_S[|w(x,t)||w_t(x,t)|+|w(x,-t)||w_t(x,-t)|]dg.\label{new}\eeq
We integrate (\ref{new}) over $(h/2,h)$ with respect to $t$ to have
\be\int_{-h/2}^{h/2}\int_S|\hat\nabla w|^2dgdt\leq\frac Ch\int_{-h}^h\int_S|w||w_t|dgdt\leq\frac Ch\|w\|\|w_t\|.\label{new1}\ee

Let
$$f(t)=|\hat\nabla w|.$$ Then
$$|f'(t)|\leq|\hat\nabla w_t|.$$
Setting $\lam=1/2,$ $a=0,$ and $b=h$ in (\ref{new2}), we obtain
\beq\int_{h/2}^h|\hat\nabla w|^2dt&&\leq4\int_0^{h/2}|\hat\nabla w|^2dt+4\int_0^h(h-t)^2|\hat\nabla w_t|^2dt\nonumber\\
&&\leq4\int_{-h/2}^{h/2}|\hat\nabla w|^2dt+4\int_{-h}^h\rho^2(t)|\hat\nabla w_t|^2dt\nonumber\eeq where
$$\rho(t)=h-t\qfq t\in(h,0);\quad\rho(t)=h+t\qfq t\in(-h,0).$$
Integrating the above inequality over $S$ with respect to the metric $g$ yields
\be\int_{h/2}^h\int_S|\hat\nabla w|^2dgdt\leq4\int_{-h/2}^{h/2}\int_S|\hat\nabla w|^2dgt+4\int_{-h}^h\int_S\rho^2(t)|\hat\nabla w_t|^2dgdt.\label{new2}\ee

On the other hand, we have
\beq0&&=\rho^2(t)w_t\hat\Delta w_t=\rho^2w_t\Delta_gw_t+\rho^2w_tw_{ttt}\nonumber\\
&&=\div_g(\rho^2w_tDw_t)+(\rho^2w_tw_{tt})_t-\rho^2|Dw_t|^2-2\rho\rho'w_tw_{tt}-\rho^2w_{tt}^2\nonumber\\
&&=\div_g(\rho^2w_tDw_t)+(\rho^2w_tw_{tt})_t-\rho^2|\hat\nabla w_t|^2-2\rho\rho'w_tw_{tt},\nonumber\eeq from which it follows that
$$\int_{-h}^h\int_S\rho^2|\hat\nabla w_t|^2dgdt=-2\int_{-h}^h\int_S\rho\rho'w_tw_{tt}dgdt\leq2\|\rho\hat\nabla w\|\|w_t\|,$$ that is,
\be\|\rho\hat\nabla w_t\|^2\leq4\|w_t\|^2.\label{new4}\ee
Using (\ref{new4}) and (\ref{new1}) in (\ref{new2}), we have
\be \int_{h/2}^h\int_S|\hat\nabla w|^2dgdt\leq\frac Ch\|w\|\|w_t\|+C\|w_t\|^2.\label{new5}\ee
A similar argument yields
\be \int_{-h}^{-h/2}\int_S|\hat\nabla w|^2dgdt\leq\frac Ch\|w\|\|w_t\|+C\|w_t\|^2.\label{new6}\ee

Finally, (\ref{2.12}) follows from (\ref{new5}), (\ref{new6}), and (\ref{new1}). \hfill$\Box$

\begin{lem}\label{newl} Let $\Upsilon(y)$ and $X(y)$ be given in $(\ref{2.6}).$ Then
$$ \Delta_gw=\div_gX(y)-\tr_g\Upsilon(y_t)+w_t\tr_g\Pi+\<DW,\Pi\>+\tr_g\ii(W)D\Pi,$$ where $\div_g$ and $\tr_g$ are the divergence and the trace of the induced metric $g,$ respectively.
\end{lem}

{\bf Proof}\,\,\,Let $x\in S$ be given. Let $E_1,$ $E_2$ be an orthonormal frame normal at $x.$ Then
$$\<E_i,E_j\>=0\quad\mbox{in a neighbourhood of $x,$}$$
$$\nabla_{E_i}\n=\lam_iE_i,\quad D_{E_i}E_j=0\qaq x\qfq 1\leq i,\,\,j\leq2,$$ where $\lam_i=\Pi(E_i,E_i)$ are the principal curvatures at $x.$ We have
\beq\Delta_gw&&=E_1\<Dw,E_1\>+E_2\<Dw,E_2\>\nonumber\\
&&=E_1\<Dw-\ii(W)\Pi+W_t,E_1\>+E_2\<Dw-\ii(W)\Pi+W_t,E_2\>\nonumber\\
&&\quad+E_1\<\ii(W)\Pi-W_t,E_1\>+E_2\<\ii(W)\Pi-W_t,E_2\>\nonumber\\
&&=\div_gX(y)+D\Pi(W,E_1,E_1)+D\Pi(W,E_2,E_2)+\Pi(D_{E_1}W,E_1)+\Pi(D_{E_2}W,E_2)\nonumber\\
&&\quad-(\<D_{E_1}W_t,E_1\>+\<D_{E_2}W_t,E_2\>+w_t\tr_g\Pi)+w_t\tr_g\Pi\nonumber\\
&&=\div_gX(y)-\tr_g\Upsilon(y_t)+w_t\tr_g\Pi+\<DW,\Pi\>+\tr_g\ii(W)D\Pi.\nonumber\eeq
\hfill$\Box$

{\bf Proof of Theorem \ref{t1}}\,\,\,Let
\be I(y)=\nabla y+tp(y),\label{2.24}\ee where $p(y)$ is given in (\ref{2.4}).

{\bf Step 1}\,\,\,Let $\hat{w}$ be the solution to problem
$$\left\{\begin{array}{l}\Delta_g \hat{w}+\hat w_{tt}=0\quad\mbox{in}\quad \Om(h),\\
\hat{w}=w\quad\mbox{on}\quad\pl\Om(h).\end{array}\right.$$ Then
\beq|w-\hat w|^2&&=|\int_{-h}^t(w_t-\hat w_t)dt|^2\leq2h\int_{-h}^h|w_t-\hat w_t|^2dt,\nonumber\eeq and thus
\be\|w-\hat{w}\|^2\leq Ch^2\|\hat\nabla(w-\hat{w})\|^2.\label{2.21}\ee
From Lemma \ref{newl}, we have
\beq|\hat\nabla(w-\hat{w})|^2&&=\div_g[(w-\hat{w})D(w-\hat{w})]+[(w-\hat w)(w_t-\hat w_t)]_t-(w-\hat{w})(\Delta_g w+w_{tt})\nonumber\\
&&=\div_g[(w-\hat{w})D(w-\hat{w})]+[(w-\hat w)(w_t-\hat w_t)]_t\nonumber\\
&&\quad-(w-\hat{w})[\div_gX-\tr_g[\Upsilon(y)]_t+w_t\tr_g\Pi+w_{tt}+\<DW,\Pi\>+\tr_g\ii(W)D\Pi]\nonumber\\
&&=\div_g[(w-\hat{w})D(w-\hat{w})]+[(w-\hat w)(w_t-\hat w_t)]_t-[(w-\hat w)w\tr_g\Pi]_t\nonumber\\
&&\quad+[(w-\hat w)\tr_g\Upsilon(y)]_t-[(w-\hat w)w_t]_t+\div_g(w-\hat w)X-\<D(w-\hat w),X\>\nonumber\\
&&\quad-(w_t-\hat w_t)\tr_g\Upsilon(y)+(w_t-\hat w_t)w\tr_g\Pi+(w_t-\hat w_t)w_t\nonumber\\
&&\quad+(w-\hat w)[\<DW,\Pi\>+\tr_g\ii(W)D\Pi].\nonumber\eeq
We integrate the above identity over the product space $S\times(-h,h)$ to have, by (\ref{2.21}),
\beq\|\hat\nabla(w-\hat{w})\|^2&&\leq\|\hat\nabla(w-\hat{w})\|(\|X(y)\|+\|\tr_g\Upsilon(y)-w\tr_g\Pi\|+\|w_t\|)\nonumber\\
&&\quad+C\|w-\hat{w}\|(\|y\|+\|DW\|)\nonumber\\
&&\leq C\|\hat\nabla(w-\hat{w})\|[\|\sym I(y)\|+h(\|y\|+\|DW\|)],\nonumber\eeq that is,
\be\|\hat\nabla(w-\hat{w})\|\leq C[\|\sym I(y)\|+h(\|y\|+\|DW\|)].\label{2.18}\ee

Using (\ref{2.18}), (\ref{2.12}), (\ref{2.21}) and (\ref{2.6}), we obtain
\beq\|Dw\|^2&&\leq C\|\hat\nabla(w-\hat{w})\|^2+C\|D\hat{w}\|^2\nonumber\\
&&\leq C\|\hat\nabla(w-\hat{w})\|^2+\frac Ch(\|\hat{w}-w\|+\|w\|)(\|\hat{w}_t-w_t\|+\|w_t\|)\nonumber\\
&&\quad+C\|\hat\nabla(\hat w-w)\|^2+C\|w_t\|^2\nonumber\\
&&\leq C\|\hat\nabla(w-\hat{w})\|^2+\frac Ch\|w\|[\|\sym I(y)\|+h(\|y\|+\|DW\|)]\nonumber\\
&&\quad+C\|\hat\nabla(w-\hat w)\|\|w_t\|+C\|w_t\|^2\nonumber\\
&&\leq C(\frac1h\|\<y,\n\>\|\|\sym I(y)\|+\|y\|^2+\|\sym I(y)\|^2+\|DW\|^2).\label{2.22}\eeq
Thus we have
\beq\|W_t\|^2&&\leq C\|Dw-\ii(W)\Pi+W_t\|^2+C\|Dw\|^2+C\|W\|^2\nonumber\\
&&\leq C(\frac1h\|\<y,\n\>\|\|\sym I(y)\|+\|y\|^2+\|\sym I(y)\|^2+\|DW\|^2).\label{2.23}\eeq

From \cite[Theorem 1.1]{CJ}, there is a constant $C>0$ such that
\be\int_S|DW|^2dg\leq C\int_S(|\sym DW|^2+|W|^2)dg\leq C\int_S(|\Upsilon(y)|^2+|y|^2)dg.\label{2.20}\ee
It follows from (\ref{2.22})--(\ref{2.20}) and (\ref{2.5})--(\ref{2.6}) that
\be\|I(y)\|^2\leq C(\frac1h\|\<y,\n\>\|\|\sym I(y)\|+\|y\|^2+\|\sym I(y)\|^2).\label{2.22*}\ee

{\bf Step 2}\,\,\,From (\ref{2.4}), we have
$$|p(y)|\leq C|\nabla y|\qfq z=x+t\n\in\Om.$$ Then
$$(1-Ch)\|\nabla y\|\leq\|I(y)\|\leq (1+Ch)\|\nabla y\|,$$
\be\|\sym \nabla y\|-Ch\|\nabla y\|\leq\|\sym I(y)\|\leq \|\sym\nabla y\|+Ch\|\nabla y\|.\label{t2.29}\ee
Thus the inequality (\ref{1.1}) follows from (\ref{2.22*}). \hfill$\Box$

\subsection{Proofs of Theorems \ref{t2} and \ref{t3} }

\hskip\parindent Let $\X(S)$ be the set of all vector fields on $S.$ For any $X,$ $Y\in\X(S),$ the curvature operator $R_{XY}$ is defined by
$$R_{XY}=-D_XD_Y+D_YD_X+D_{[X,Y]},$$ where $[\cdot,\cdot]$ is the Lie product. The Ricci identity reads
\be D^2T(\cdots,X,Y)=D^2T(\cdots,Y,X)+R_{XY}(T)(\cdots),\label{2.28}\ee
where $T$ is a k-order tensor field. This formula can help us to exchange the order of the second-order covariant differential of a k-order tensor field.

Let $x\in S$ be given and let $e_1,$ $e_2$ be an orthonormal basis of  $M_x$ with the positive  orientation in the induced metric
$g.$
For any $W\in H^1(S,\X(S)),$ we denote a 2-form $\si(W)$ on $S$ by
$$\si(W)=D_{e_1}W\wedge_gD_{e_2}W\quad\mbox{at}\quad x,$$ where $\wedge_g$ is the exterior product of the induced metric $g$ on $S.$ Then $\si(W)$ is well defined. In fact, let $\hat e_1,$ $\hat e_2$ be another  orthonormal basis with the positive  orientation. Suppose that
$$e_1=\a_{11}\hat e_1+\a_{12}\hat e_2,\quad e_2=\a_{21}\hat e_1+\a_{22}\hat e_2.$$ Then
$$(\a_{ij})(\a_{ij})^T=I,\quad \det(\a_{ij})=1£¬$$ where $I$ is the identity matrix in $\R^2.$ It follows that
\beq D_{e_1}W\wedge_gD_{e_2}W&&=(\a_{11}D_{\hat e_1}W+\a_{12}D_{\hat e_2}W)\wedge_g(\a_{21}D_{\hat e_1}W+\a_{22}D_{\hat e_2}W)\nonumber\\
&&=(\a_{11}\a_{22}-\a_{12}\a_{21})D_{\hat e_1}W\wedge_gD_{\hat e_2}W=D_{\hat e_1}W\wedge_gD_{\hat e_2}W.\nonumber\eeq
Then there is a function $\var$ on $S,$ independent of the choice of orthonormal base, such that
\be\si(W)=\var(x)\E\qfq x\in S,\label{2.24}\ee where $\E$ is the volume element of the induced metric $g.$

\begin{lem}\label{ll2.5} For any $W\in H^1(S,\X(S)),$ we have
\be2\int_S\var dg=\int_S\kappa|W|^2dg+\int_{\pl S}[2\<W,\mu\>\tau\<W,\tau\>+\<D_\tau\mu,\tau\>|W|^2]d\pl S,\label{2.30}\ee
where $\var$ is given in $(\ref{2.24})$ and $\mu$ and $\tau$ are the outside normal and the tangential along the boundary $\pl S$ in the induced metric $g,$ respectively.
\end{lem}

{\bf Proof}\,\,\,For $W$ given, we denote a vector field $B(W)$ on $S$ by
\be B(W)=(W\wedge_g\ii(e_2)D^TW)(e_1,e_2)e_1-(W\wedge_g\ii(e_1)D^TW)(e_1,e_2)e_2\qfq x\in S,\label{2.29}\ee where $e_1,$ $e_2$ is an orthonormal basis of $M_x$ and $D^TW$ is the transpose of $DW.$
It is easy to check that the definition of $B(W)$ is independent of the choice of $e_1,$ $e_2.$

Since $D_\tau\mu=\<D_\tau\mu,\tau\>\tau$ and $D_\tau\tau=-\<D_\tau\mu,\tau\>\mu$ on the boundary $\pl S,$ we have
\beq\<B(W),\mu\>&&=(W\wedge_g\ii(\tau)D^TW)(\mu,\tau)=\<W,\mu\>\<D_\tau W,\tau\>-\<W,\tau\>\<D_\tau W,\mu\>\nonumber\\
&&=\<W,\mu\>\tau\<W,\tau\>-\<W,\tau\>\tau\<W,\mu\>-\<W,\mu\>\<W,D_\tau\tau\>+\<W,\tau\>\<W,D_\tau\mu\>\nonumber\\
&&=2\<W,\mu\>\tau\<W,\tau\>-\tau(\<W,\mu\>\<W,\tau\>)+|W|^2\<D_\tau\mu,\tau\>.\label{2.26}\eeq

Let $x\in S$ be given. Let $E_1,$ $E_2$ be a frame field normal at $x$  with the positive  orientation. Then
$$D_{E_i}E_j=0\quad\mbox{at}\quad x\qfq1\leq i,\,\,j\leq2.$$
It follows (\ref{2.24}), (\ref{2.28}), and (\ref{2.29}) that
\beq \var(x)&&=\si(W)(E_1,E_2)=\<D_{E_1}W,E_1\>\<D_{E_2}W,E_2\>-\<D_{E_2}W,E_1\>\<D_{E_1}W,E_2\>\label{2.27}\\
&&=E_1(\<W,E_1\>\<D_{E_2}W,E_2\>)-\<W,E_1\>D^2W(E_2,E_2,E_1)\nonumber\\
&&\quad-E_2(\<W,E_1\>\<D_{E_1}W,E_2\>)+\<W,E_1\>D^2W(E_2,E_1,E_2)\nonumber\\
&&=E_1(\<W,E_1\>\<D_{E_2}W,E_2\>-\<W,E_2\>\<D_{E_2}W,E_1\>)+E_1(\<W,E_2\>\<D_{E_2}W,E_1\>)\nonumber\\
&&\quad+E_2(-\<W,E_1\>\<D_{E_1}W,E_2\>+\<W,E_2\>\<D_{E_1}W,E_1\>)-E_2(\<W,E_2\>\<D_{E_1}W,E_1\>)+\kappa\<W,E_1\>^2\nonumber\\
&&=\div_g B(W)+\<D_{E_1}W,E_2\>\<D_{E_2}W,E_1\>-\<D_{E_2}W,E_2\>\<D_{E_1}W,E_1\>\nonumber\\
&&\quad+\<W,E_2\>[D^2W(E_1,E_2,E_1)-D^2W(E_1,E_1,E_2)]+\kappa\<W,E_1\>^2\nonumber\\
&&=\div_gB(W)-\var(x)+\kappa|W|^2,\label{2.31}\eeq where the formula (\ref{2.27}) has been used.
Thus (\ref{2.30}) follows from (\ref{2.31}) and (\ref{2.26}). \hfill$\Box$\\

In the sequel, for a vector field $W\in\X(S),$ we denote
$$W_i=\<W,E_i\>,\quad W_{ij}=DW(E_i,E_j)=\<D_{E_j}W,E_i\>\qfq 1\leq i,\,\,j\leq2,$$ where $E_1,$ $E_2$ is an orthonormal frame on $S.$ From (\ref{2.27}), we have
\be \var(x)=W_{11}W_{22}-W_{12}W_{21},\label{2.32}\ee
where $\var$ is given in (\ref{2.24}). Moreover, if $f$ is a function, we denote
$$W(f)=\<W,Df\>.$$

We need the following.

\begin{lem}\label{l1.6} Let $M$ be of $\CC^3.$ Let  $\lam(q)$ be a principal curvture for each $q\in M.$ Let $p\in M$ be given.
Suppose that there is a neighbourhood $\N$ of $p$
such that the following assumptions hold.

$(i)$\,\,\,$\lam\in \CC^1(\N);$

$(ii)$\,\,the algebraic multiplicity of $\lam(q)=$ the  geometric multiplicity $=1$ for all $q\in\N.$

Then there exists locally a $\CC^1$ vector field $X$ such that
$$ \nabla_X\n=\lam X\quad\mbox{in a neighbourhood of $p.$}$$
\end{lem}

{\bf Proof}\,\,\,Let $\psi:$ $\N\rw\R^2$ be a local coordinate at $p$ with $\psi(q)=(x_1,x_2)$ and $\psi(p)=0.$ Consider the matrices
$$A(x)=\Big(a_{ij}(x)\Big),\quad \nabla_{\pl x_i}\n=a_{1i}(x)\pl x_1+a_{2i}(x)\pl x_2\qfq1\leq i,\,\,j\leq2.$$
From (ii)
$$\rank\Big(\lam(x)\delta_{ij}-a_{ij}(x)\Big)=1\qfq x\quad\mbox{in a neighbourhood of $0.$}$$
We may assume that
$$\Big(\lam(0)-a_{11}(0),\,\,-a_{12}(0)\Big)\not=0.$$
Thus
$$\Big(\lam(x)-a_{11}(x),\,\,-a_{12}(x)\Big)\not=0\qfq x\quad\mbox{in a neighbourhood of $0.$}$$
Let
$$X=a_{12}(x)\pl x_1+[\lam(x)-a_{11}(x)]\pl x_2.$$
Obviously, the above $X$ meets our need.\hfill$\Box$\\

For each $p\in M,$ we denote by $Q:$ $M_p\rw M_p$ the rotation by $\pi/2$ along the clockwise direction, which is very useful in the case of the negative curvature, see  \cite{Yao2017}.
For any $\a\in M_p,$ $\a,$ $Q\a$ forms an orthonormal basis on $M_p.$

\begin{pro}\label{pr2.1} Let $p\in M$ be given. Suppose that there are two different principal curvatures, $\lam_1\not=\lam_2,$ at $p.$ Then  there exists a local principal coordinate
$\psi=x$ around $p,$ i.e.,
$$\nabla_{\pl x_i}\n=\lam_i\pl x_i\quad\mbox{in a neighbourhood of $p$}\qfq i=1,\,\,2.$$
\end{pro}

{\bf Proof}\,\,\, From Lemma \ref{l1.6} there is a vector field $X$ with $|X|=1$
such that
\be\nabla_X\n=\lam_1 X\quad\mbox{in a neighbourhood of $p$}.\label{6.33}\ee Let $Y=QX.$ Then $X,$ $QX$ forms an orthonormal basis. Thus
\be\nabla_Y\n=\lam_2Y\quad\mbox{in a neighbourhood of $p$}.\label{6.34}\ee

We claim there exist functions $f_1$ and $f_2$ such that
\be[f_1X,f_2Y]=0.\label{5.33}\ee
We define a curve by
$$\a'(t)=X(\a(t))\qfq t\in(-\varepsilon,\varepsilon),\quad \a(0)=p.$$
Then for $t\in(-\varepsilon,\varepsilon)$ given, we solve problem
\be\b_s(t,s)=Y(\b(t,s))\qfq s\in(-\varepsilon_1,\varepsilon_1),\quad\b(t,0)=\a(t).\label{4.23}\ee
Since
$$\det\Big(\b_t(0,0),\b_s(0,0)\Big)=\det\Big(X(p),Y(p)\Big)=\pm1\not=0,$$
the map $\psi(\b(t,s))=(t,s)$ forms a local coordinate at $p$ with (\ref{4.23}) true.
We let
$$f_1(\b(t,s))=e^{\int_{-\varepsilon_1}^s\<D_XY,X\>(\b(t,s))ds}\qfq (t,s)\in(-\varepsilon,\varepsilon)\times(-\varepsilon_1,\varepsilon_1).$$
Then $f_1$ satisfies
\be Y(f_1)=f_1\<D_XY,X\>.\label{4.35}\ee Similarly, there is a function $f_2$ such that
\be X(f_2)=f_2\<D_YX,Y\>.\label{4.36}\ee
(\ref{5.33}) follows from (\ref{4.35}) and (\ref{4.36}).

Let $\a(t,p)$ and $\b(t,p)$ be the flows by $f_1X$ and $f_2Y$ at a neighbourhood of $\{0\}\times\{p\},$ respectively. From \cite[p.233, Theorem 9.44]{Lee}, we have
$$\a(t,\b(s,q))=\b(s,\a(t,q)).$$
Then (\ref{5.33}) implies that $\hat\psi(\a(t,\b(s,p)))=(t,s)$ is a local coordinate such that
$$\pl t=f_1X,\quad \pl s=f_2Y.$$ \hfill$\Box$

Next, we consider a rigidity lemma on the strain tensor of the middle surface. In the case of the parabolic or the hyperbolic, it has established in
\cite{GH}-\cite{Ha2} when the middle surface is given by a single principal coordinate. In the case of the elliptic shell, it has been given in \cite{CL} if
the middle surface consists of  a single coordinate. Here we treat it coordinates free, which particularly includes the case of the closed elliptic shells.

\begin{pro}\label{pr2.2} Suppose $\Om$ is a parabolic shell. Then there is $C>0$ such that
\be \|W\|^2_{L^2(S)}\leq C\|\Upsilon(y)\|_{L^2(S)}(\|\Upsilon(y)\|_{L^2(S)}+\|w\|_{L^2(S)})\label{2.37}\ee for any $y=W+w\n\in H_0^1(S,\R^3).$
\end{pro}

{\bf Proof}\,\,\,Let $\hat S$ be a bounded open region on $M$ such that
\be\overline{S}\subset\hat S;\quad \kappa(x)=0,\quad\nabla\n\not=0\qfq x\in\hat S.\label{2.38}\ee
For $y\in H_0^1(S,\R^3),$ we extend $y\in H_0^1(\hat S,\R^3)$ by
$$y=0\qfq x\in\hat S/S.$$ In the above sense, we have
$$H_0^1(S,\R^3)\subset H_0^1(\hat S,\R^3).$$
Thus (\ref{2.37}) follows from Lemma \ref{l2.7} below. \hfill$\Box$

\begin{lem}\label{l2.7} Let $\hat S\subset M$ be such that $(\ref{2.38})$ hold. Let $p\in\hat S$ be given and $\gamma>0$ be given small. Then exist a neighbourhood $\N$ of $p$ and  constants
$C>0,$ independent of $\gamma,$ and $C_\gamma>0,$
such that
\be\|W\|_{L^2(\N)}^2\leq C\gamma\|W\|^2_{L^2(\hat S)}+C_\gamma\|\Upsilon(y)\|_{L^2(\hat S)}(\|\Upsilon(y)\|_{L^2(\hat S)}+\|w\|_{L^2(\hat S)})\label{2.39}\ee for any $y=W+w\n\in H_0^1(\hat S,\R^3).$
\end{lem}

{\bf Proof}\,\,\,From Lemma \ref{l1.6} there is a vector field $X$ with $|X|=1$
such that (\ref{6.33}) and (\ref{6.34}) hold for $x$ in a neighbourhood of $p,$ where $\lam_1=\tr_g\Pi,$ $\lam_2=0,$ and $Y=QX.$ It follows from (\ref{6.34}) that
\be\nabla_YX=D_YX=aY,\quad \nabla_YY=D_YY=\<D_YY,X\>X=-aX,\label{2.42}\ee where $a=\<D_YX,Y\>.$

Let $\a(\cdot):$ $(-\varepsilon,\varepsilon)\rw\hat S$ be the curve with
$$\a(0)=x,\quad \a'(t)=X(\a(t))\qfq t\in(-\varepsilon,\varepsilon).$$
Then we define $\b:$ $(-\varepsilon,\varepsilon)\times(-\varepsilon_1,\varepsilon_1)$ by
$$\b_s(t,s)=Y(\b(t,s))\qfq (t,s)\in(-\varepsilon,\varepsilon)\times(-\varepsilon_1,\varepsilon_1);\quad \b(t,0)=\a(t)\qfq t\in(-\varepsilon,\varepsilon).$$
Since
$$\det\Big(\b_t(0,0),\b_s(0,0)\Big)=\det\Big(X(p),QX(p)\Big)=\pm1\not=0,$$ the map $\psi(\b(t,s))=(t,s)$ forms a coordinate at $p.$ We set
$$\N=\{\,\b(t,s)\,|\,(t,s)\in(-\varepsilon,\varepsilon)\times(-\varepsilon_1,\varepsilon_1)\,\},$$ where $\varepsilon>0$ and $\varepsilon_1>0$ are small enough.

{\bf Step 1}\,\,\,We claim that, for each $t\in(-\varepsilon,\varepsilon)$ fixed,

(1)\,\,\,the curve $\b(t,\cdot)$ has no self-intersection point for $s\in(-\varepsilon_1,\varepsilon_1);$

(2)\,\,\, the vector fields $X$ and $Y$ and the curve $\b(t,\cdot)$ can be simultaneously extended to outside of $\hat{S}$ from both  directions, i.e., there are $s_-(t)<0$ and $s_+(t)>0$ satisfying
$$\b(t,s_\pm(t))\in\pl\hat{S}.$$

For convenience, we denote $\b(s)=\b(t,s).$ Let
$$\b(s)=\b_1(s)X+\b_2(s)Y+b_3(s)\n\qfq s\in(-\varepsilon_1,\varepsilon_1).$$
Using (\ref{6.34}) and (\ref{2.42}), we have
\beq\b'(s)&&=\b_1'(s)X+\b_2'(s)Y+\b_3'(s)\n+\b_1(s)\nabla_YX+\b_2(s)\nabla_YY+\b_3(s)\nabla_Y\n\nonumber\\
&&=[\b_1'(s)-a\b_2(s)]X+[\b_2'(s)+a\b_1(s)]Y+\b_3'(s)\n,\nonumber\eeq which yields, since $\b'(s)=Y,$
\be\b_1'(s)-a\b_2(s)=0,\quad \b_2'(s)+a\b_1(s)=1,\quad\b_3'(s)=0.\label{2.43}\ee

On the other hand, using the formula
$$\nabla_X\nabla_Y\n=\nabla_Y\nabla_X\n+\nabla_{[X,Y]}\n,$$
and from (\ref{6.33}) and (\ref{6.34}), we obtain
$$ [Y(\lam_1)+\lam_1\<[X,Y],X\>]X+\lam_1 aY=0,$$ that is, $a=0,$ since $\lam_1\not=0.$ It follows from (\ref{2.43}) that
$$\b(s)=\b_1(0)X+[\b_2(0)+s]Y+\b_3(0)\n\qfq s\in(-\varepsilon_1,\varepsilon_1),$$ which proves (1) and (2) by Lemma \ref{l1.6}.

{\bf Step 2}\,\,\,Let $\var$ be given in (\ref{2.24}). From (\ref{6.33}), (\ref{6.34}), and (\ref{2.32}), we have
\beq|\Upsilon(y)|^2&&=[DW(X,X)+\lam w]^2+\frac12[DW(X,Y)+DW(Y,X)]^2+[DW(Y,Y)]^2\nonumber\\
&&\geq\frac12\{[DW(X,Y)]^2+[DW(Y,X)]^2\}-\var+DW(X,X)DW(Y,Y)\nonumber\\
&&=\frac12\{[DW(X,Y)]^2+[DW(Y,X)]^2\}-\var+[\Upsilon(y)(X,X)-\lam w]\Upsilon(y)(Y,Y),\nonumber\eeq that is,
\be [DW(X,Y)]^2+[DW(Y,X)]^2\leq C|\Upsilon(y)|(|\Upsilon(y)|+|w|)+2\var.\label{3.34} \ee

{\bf Step 3}\,\,\,For $t\in(-\varepsilon,\varepsilon)$ given, from Step 1, we have
\beq|W|^2&&=2\int_{s_-(t)}^s\<D_YW,W\>ds=2\int_{s_-(t)}^s[\<W,X\>DW(X,Y)+\<W,Y\>DW(Y,Y)]ds\nonumber\\
&&\leq\gamma\int_{s_-(t)}^{s_+(t)}|W|^2ds+C_\gamma\int_{s_-(t)}^{s_+(t)}\{[DW(X,Y)]^2+|\Upsilon(y)|^2\}ds,\nonumber\eeq for $\gamma>0$ small.
We integrate the above inequality in $(t,s)$ over $(-\varepsilon,\varepsilon)\times(-\varepsilon_1,\varepsilon_1)$ to have, by (\ref{3.34}) and (\ref{2.30}),
\beq \int_\N|W|^2dg&&\leq\gamma\|W\|^2_{L^2(\hat S)}+C_\gamma\int_{-\varepsilon}^\varepsilon\int_{s_-(t)}^{s_+(t)}\{[DW(X,Y)]^2+|\Upsilon(y)|^2\}dsdt\nonumber\\
&&\leq\gamma C\|W\|^2_{L^2(\hat S)}+C_\gamma\int_{-\varepsilon}^\varepsilon\int_{s_-(t)}^{s_+(t)}\{|\Upsilon(y)|(|\Upsilon(y)|+|w|)+2\var+|\Upsilon(y)|^2\}dsdt\nonumber\\
&&\leq\gamma C\|W\|^2_{L^2(\hat S)}+C_\gamma\int_{\hat S}\{C|\Upsilon(y)|(|\Upsilon(y)|+|w|)+2\var+|\Upsilon(y)|^2\}dg\nonumber\\
&&\leq\gamma C\|W\|^2_{L^2(\hat S)}+C_\gamma\|\Upsilon(y)\|_{L^2(\hat S)}(\|\Upsilon(y)\|_{L^2(\hat S)}+\|w\|_{L^2(\hat S)}).\nonumber\eeq
The proof is complete.  \hfill$\Box$

\begin{pro}\label{pr2.3} Let $S$ be elliptic.  There is $C>0$ such that
\be \|DW\|^2_{L^2(S)}+\|w\|^2_{L^2(S)}\leq C\|\Upsilon(y)\|_{L^2(S)}^2+C|\int_S\var dg|\label{3.44}\ee for any $y=W+w\n\in H^1(S,\R^3).$
\end{pro}

It follows from Proposition \ref{pr2.3} immediately that

\begin{cor}\label{p2.3} Let  $S$ be elliptic.

$(i)$\,\,\, If $|\pl S|>0,$  then there is $C>0$ such that
\be \|y\|^2_{L^2(S)}\leq C\|\Upsilon(y)\|_{L^2(S)}^2\label{2.44}\ee for any $y=W+w\n\in H^1_0(S,\R^3).$

$(ii)$\,\,\,If  $S$ is a closed surface, then there is $C>0$ such that, for  any $y=W+w\n\in H^1(S,\R^3),$ there exists an infinitesimal identity $y_0\in H^1(S,\R^3),$
satisfying
\be \|y-y_0\|^2_{L^2(S)}\leq C\|\Upsilon(y)\|_{L^2(S)}^2.\label{2.44}\ee
\end{cor}

{\bf Proof of Proposition \ref{pr2.3}}\,\,\,Let $p\in S$ be given. Let $e_1,$ $e_2$ be an orthonormal basis of $M_p$ with the positive orientation such that
$$\nabla_{e_i}\n=\lam_ie_i\qaq p\qfq1\leq i,\,\,j\leq2.$$ Let $E_1,$ $E_2$ be a frame field normal at $p$ such that
$$E_1(p)=e_1,\quad E_2(p)=e_2.$$ Then
$$\<E_i,E_j\>=\delta_{ij} \quad\mbox{a neighbourhood of $p,$}$$ and
$$D_{E_j}E_i=0\qaq p.$$

Using the above formulas, we compute at $p,$ for $\varepsilon>0$ and $\varsigma>0$ small,
\beq|\Upsilon(y)|^2&&=|W_{11}+\lam_1w|^2+\frac12|W_{12}+W_{21}|^2 +|W_{22}+\lam_2w|^2\nonumber\\
&&=W_{11}^2+W_{22}^2+\frac12|W_{12}+W_{21}|^2 +2(\lam_1W_{11}+\lam_2W_{22})w+|\Pi|^2w^2\nonumber\\
&&\geq W_{11}^2+W_{22}^2+\frac12|W_{12}+W_{21}|^2 -\frac{1}{|\Pi|^2-\varepsilon}(\lam_1W_{11}+\lam_2W_{22})^2+\varepsilon w^2\nonumber\\
&&=\frac{1}{|\Pi|^2-\varepsilon}[(\lam_2^2-\varepsilon) W_{11}^2-2\kappa W_{11}W_{22}+(\lam_1^2-\varepsilon)W_{22}^2]+\frac12|W_{12}+W_{21}|^2+\varepsilon w^2\nonumber\\
&&\geq\frac{1}{|\Pi|^2-\varepsilon}\{(\lam_2^2-\varepsilon) W_{11}^2-2[\kappa-(|\Pi|^2-\varepsilon)\varsigma] W_{11}W_{22}+(\lam_1^2-\varepsilon)W_{22}^2\}\nonumber\\
&&\quad+\varsigma(W_{12}^2+W_{21}^2)-2\varsigma\var(p)+\varepsilon w^2\nonumber\\
&&=\frac{1}{|\Pi|^2-\varepsilon}\{\si(|W_{11}|^2+|W_{22}|^2)+(\sqrt{\lam_2^2-\varepsilon-\si} W_{11}-\sqrt{\lam_1^2-\varepsilon-\si}W_{22})^2\}\nonumber\\
&&\quad+\varsigma(W_{12}^2+W_{21}^2)-2\varsigma\var(p)+\varepsilon w^2,\label{2.47}\eeq
where $W_{ij}=DW(E_i,E_j),$ $\var$ is given in (\ref{2.32}), and $\si>0$ is given through the formula
$$(\lam_2^2-\varepsilon-\si)(\lam_1^2-\varepsilon-\si)=[\kappa-(|\Pi|^2-\varepsilon)\varsigma]^2,$$ when $\varepsilon>0$ and $\varsigma>0$ are small enough.

We integrate (\ref{2.47}) over $S$ to obtain (\ref{3.44}) from Lemma \ref{ll2.5}. \hfill$\Box$\\

{\bf Proof of Theorem \ref{t2} }\,\,\, We use the norm (\ref{b3}) and the formulas (\ref{2.5}) and (\ref{2.6}). Let $\lam=\tr_g\Pi>0$ for $x\in\overline{S}.$ Let
$I(y)=\nabla y+tp(y)$ where $p(y)$ is given in (\ref{2.4}). We integrate (\ref{2.37}) in $t\in(-h,h)$ to obtain, by (\ref{2.6}),
\be \|W\|^2\leq C(\g\varepsilon^2\|w\|^2+\frac{1}{\g\varepsilon^2}\|\sym I(y)\|^2),\label{t2.53}\ee  for any $1>\varepsilon>0$ and $1>\gamma>0$ small.
In addition, we have
\beq w^2\lam&&=w\tr_g(w\Pi)=w\tr_g\Upsilon(y)-w\div_gW=w\tr_g\Upsilon(y)-\div_g(wW)+\<Dw,W\>,\nonumber\eeq which yield, by (\ref{2.5}), (\ref{2.6}), and (\ref{t2.53}),
\beq \|w\|^2_{L^2(S)}&&\leq C(\|\Upsilon(y)\|_{L^2(S)}^2+\|Dw\|_{L^2(S)}\|W\|_{L^2(S)} )\nonumber\\
&&\leq C(\|\Upsilon(y)\|_{L^2(S)}^2+\|I(y)\|_{L^2(S)}\|W\|_{L^2(S)}+\|W\|^2_{L^2(S)} )\nonumber\\
&&\leq C(\|\sym I(y)\|_{L^2(S)}^2+\varepsilon^2\|I(y)\|_{L^2(S)}+\frac1{\varepsilon^2}\|W\|^2_{L^2(S)} ) \nonumber\\
&&\leq C(\varepsilon^2\|I(y)\|_{L^2(S)}+\g\|w\|^2_{L^2(S)}+\frac1{\g\varepsilon^4}\|\sym I(y)\|^2_{L^2(S)} ),\label{t2.54}\eeq
where $1>\varepsilon>0$ is given small.
It follows from (\ref{t2.54}) that
\be \|w\|^2\leq C_\gamma(\varepsilon^2\|I(y)\|^2+\frac1{\varepsilon^4}\|\sym I(y)\|^2),\label{t2.55}\ee for any $1>\varepsilon>0$ small.  Thus we have
\be \|w\|\leq C(\varepsilon\|I(y)\|+\frac1{\varepsilon^2}\|\sym I(y)\|). \label{t2.56}\ee Next inserting (\ref{t2.55}) into (\ref{t2.53}) gives
\be \|W\|^2\leq C(\varepsilon^4\|I(y)\|^2+\frac1{\varepsilon^2}\|\sym I(y)\|^2).\label{t2.57}\ee It follows from (\ref{t2.55}) and (\ref{t2.57}) that
\be\|y\|^2\leq C(\varepsilon^2\|I(y)\|^2+\frac1{\varepsilon^4}\|\sym I(y)\|^2),\label{t2.58}\ee for any $1>\varepsilon>0$ small.

We now let $\varepsilon=h^{1/4}$ in (\ref{t2.56}) and (\ref{t2.58}). From (\ref{t2.56}), we have
\beq \frac{1}{h}\|w\|\|\sym I(y)\|&&\leq C\|I(y)\|\frac{\|\sym I(y)\|}{h^{3/4}}+C\frac{\|\sym I(y)\|^2}{h^{3/2}}\nonumber\\
&&\leq C\g\|I(y)\|^2+C_\gamma\frac{\|\sym I(y)\|^2}{h^{3/2}}, \label{t2.59}\eeq for any $\gamma>0$ small. Using (\ref{t2.59}) and (\ref{t2.58}) with $\varepsilon=h^{1/4}$ in (\ref{2.22*}), we obtain (\ref{1.3}) from (\ref{t2.29}).
\hfill$\Box$\\

{\bf Proof of Theorem \ref{t3}}\,\,\,Estimate (\ref{1.3}) follows from the inequality (\ref{2.44}), Lemma \ref{lem2.1}, and Theorem \ref{t1}.
\hfill$\Box$

\subsection{Proof Theorem \ref{t4};\,Ansatz}
\hskip\parindent Here we use the norm (\ref{b3}).

(i)\,\,\,{\bf Let $\Om$ be parabolic.} From Proposition  \ref{pr2.1}, a local principal coordinate exists on $S.$ In such a principal coordinate an ansatz has been constructed in
\cite[Theorem 3.3]{GH}.

(ii)\,\,\,{\bf Let $\Om$ be elliptic.} Set
$$\kappa_0=\sup_{p\in S}\kappa(p).$$

Let $p_0\in S$ be given and let $\si_0>0$ be such that
$$\overline{{\bf B}(p_0,\si_0)}\subset S,\quad \frac{\sin\sqrt{\kappa_0}t}{\sqrt{\kappa_0}t}\geq\frac12\qfq t\in[0,\si_0],$$ where ${\bf B}(p_0,\si_0)$ is the geodesic plate in the induced metric $g$ centered at $p_0$ with radius $\si_0.$ Let $\var\in C^2_0(S)$ be such that
$$\var(p)=1\qfq p\in{\bf B}(p_0,\si_0).$$ Let
$\rho(p)=d_g(p,p_0)$ be the distance from $p\in S$ to $p_0$ in the induced metric $g$ on $M.$ We set
$$y=W+w\n,\quad w=\var\cos(\phi\rho),\quad W=-tDw,\quad \phi=\frac{1}{h^{1/2}}.$$
Denote ${\bf B}(\si_0)$ by the plate in $M_{p_0}$ centered at the origin with radius $\si_0.$ Let $dx$ be the volume element in $M_{p_0}.$
From the volume comparison theorem, we have
\beq\int_Sw^2dg&&\geq\int_{{\bf B}(p_0,\si_0)}\cos^2(\phi\rho)dg\geq\int_{|x|<\si_0}\cos^2(\phi|x|)\frac{\sin(\sqrt{\kappa_0}|x|)}{\sqrt{\kappa_0}|x|}dx\geq\frac12\int_{|x|<\si_0}\cos^2(\phi|x|)dx\nonumber\\
&&=\pi\int_0^{\si_0}r\cos^2(\phi r)dr=\frac{\pi}{2}\int_0^{\si_0}r[1-\cos(2\phi r)]dr\geq\frac{\pi}{2}\sum_{k=0}^m\int_{(k+\frac14)h^{1/2}\pi}^{(k+\frac34)h^{1/2}\pi}r[1-\cos(2\phi r)]dr\nonumber\\
&&\geq\frac{(m+1)^2h\pi^3}8\geq\frac{\pi^3}8(\frac{\si_0}{\pi}-\frac{3h^{1/2}}4)^2,\label{2.48}\eeq where
$$m=[\frac{\si_0}{h^{1/2}\pi}-\frac{3}{4}].$$

Moreover, we have
$$Dw=-\phi\sin(\phi\rho)D\rho,\quad D^2w=-\phi^2\cos(\phi\rho)D\rho\otimes D\rho-\phi\sin(\phi\rho)D^2\rho,$$ that yield
\be |Dw|^2\leq\frac Ch,\quad |D^2w|^2\leq \frac C{h^2}\qfq p\in S.  \ee
Noting that $|D\rho|=1,$ by a similar computation as in (\ref{2.48}), we obtain
\be \frac{\si_1}{h}\leq\int_S|Dw|^2dg\leq \frac Ch.\ee
In addition, a simple computation shows that
\be \|\nabla y+tp(y)\|^2=2h\int_S(w^2|\Pi|^2+2|Dw|^2)dg+\frac{2h^3}{3}\int_S(|D^2w|^2+|\ii(Dw)\Pi|^2)dg,\ee
\be \|\sym\nabla y+t\sym p(y)\|^2=2h\int_Sw^2|\Pi|^2dg+\frac{2h^3}{3}\int_S(\frac14|\ii(Dw)\Pi|^2+|D^2w|^2)dg.\label{2.52}\ee

Finally, it follows from (\ref{2.48})-(\ref{2.52}) that
$$\frac{\|\nabla y\|^2}{\|\sym\nabla y\|^2} \sim\frac{C}h.$$\hfill$\Box$

{\bf Conflict of interest statement}

There is no conflict of interests.

Ethical approval: This article does not contain any studies with human participants or animals performed by the author.

 \end{document}